\documentclass[%
 reprint,
 amsmath,amssymb,
 aps,
]{revtex4-2}

\usepackage{color}
\usepackage{float}
\usepackage{graphicx}
\usepackage{dcolumn}
\usepackage{bm}

\begin{document}

\preprint{APS/123-QED}

\title{Scattering in Time-Varying Drude-Lorentz Models}

\author{Bryce Dixon}
 \email{bd369@exeter.ac.uk}
\author{Calvin M. Hooper}%
\author{Ian R. Hooper}
\author{Simon A. R. Horsley}
\affiliation{%
 Department of Physics and Astronomy, University of Exeter, \\ Stocker Road, Exeter, EX4 4QL, UK 
}%

\date{\today}

\begin{abstract}
Motivated by recent experiments, the theoretical study of wave propagation in time varying materials is of current interest.  Although significant in nearly all such experiments, material dispersion is commonly neglected in theoretical studies.  Yet, as we show here, understanding the precise microscopic model for the material dispersion is crucial for predicting experimental outcomes.  Here we study the temporal scattering coefficients of four different time-varying Drude-Lorentz models, exploring how an incident continuous wave splits into forward and backward waves due to an abrupt change in plasma frequency. The differences in the predicted scattering are unique to time-varying media, and arise from the exact way in which the time variation appears in the various model parameters.  We verify our results using a custom finite difference time domain algorithm, concluding with a discussion of the limitations that arise from using these models with an abrupt change in plasma frequency.
\end{abstract}

\maketitle


\section{\label{intro} Introduction}

The potential for manipulating the frequency spectrum of a wave, to the degree already achieved with spatially structured materials, has motivated recent interest in time-varying media within both photonics and metamaterials communities~\cite{Galiffi_2022}.  Despite the desire to translate spatial concepts into temporal ones, this field is not simply a replacement $x\to t$ of well established results.  For an adiabatic variation of e.g. the refractive index, an incident wave undergoes `time refraction'~\cite{Zhou2020, Bohn:21}, its frequency shifting with the refractive index, whereas a rapid temporal switch of material parameters leads to `time reflection'~\cite{1124533, J_T_Mendonça_2002, Bacot2016, Moussa_2023, Xiao:14} (and even the temporal analogue of an  anti--reflection `coating'~\cite{Pacheco-Pena:20}).  In the latter effect, the constraint of causality requires there is no wave propagating backwards in time, which might be expected from a naive analogy with reflection from a spatial interface.  Instead the `reflected' wave propagates forwards in time, but evolves as its time reverse, retracing its motion back to its source (a novel effect, observed in~\cite{Bacot2016,Moussa_2023}).  Similarly, a periodic modulation of material properties in time leads to a band structure, just as for spatially periodic media.  Yet unlike spatially periodic media, where a band gap implies an exponential decay of a propagating wave, a band gap in a temporally modulated medium indicates the onset of parametric amplification, with the wave amplitude growing exponentially in time~\cite{1138637, 8434236, Koutserimpas:22}.  Topological effects arise in these modulated materials, edge states translating into signals that abruptly transition from being amplified to being absorbed~\cite{Lustig:18}.

If the material is anisotropic and time--modulated, additional possibilities arise which have only a tenuous analogue in spatially inhomogeneous materials, such as `temporal aiming'~\cite{Pacheco-Peña2020}, where the direction of propagation is shifted without any spatial boundaries, or the `temporal Brewster angle'~\cite{PhysRevB.104.214308}, where time reflection vanishes for a particular transition in the material anisotropy.   We note that one motivation for this field of research is the possibility of achieving non-reciprocal effects without magnetism~\cite{Sounas2017}.  

Much existing theory, particularly calculations of `reflection' and `transmission' at a temporal interface, is restricted to non-dispersive materials, where the medium is assumed to react instantaneously to the electromagnetic field.  However, this is rarely a good approximation, especially in the case of a temporal interface.  For example, at near infrared wavelengths a temporal interface implies switching on a femtosecond timescale.  Neglecting dispersion is thus equivalent to assuming a material response time that is sub--femtosecond.  Yet typical materials such as Indium Tin Oxide have a collision rate on the order of $~200{\,\rm THz}$, which implies a decay time of the material response of around $5\,{\rm fs}$.  In this example (which is based on the work presented in e.g.~\cite{kinsey2015,Jaffray:22, Bohn2021,tirole2022,tirole2023}) it thus seems impossible to neglect dispersive effects.

Assuming a Lorentzian model for the material dispersion, it has been shown that~\cite{solís2021timevaryingmaterialspresencedispersion}, through conservation of wave--vector $\boldsymbol{k}$, \emph{four} modes couple at a temporal interface, leading to two forward (positive frequency) and two backward (negative frequency) propagating waves, taking the form of a high and low frequency pair.  Unlike the non--dispersive case, where the refractive index is simply a number that we can choose to depend on time, in such a dispersive system we must make some aspect of the material dynamics as time dependent (e.g. the collision rate, the damping constant, the resonant frequency, or the field--medium coupling constant).  In \cite{solís2021timevaryingmaterialspresencedispersion}, the plasma frequency was taken to vary abruptly, but to model any particular experiment it is far from obvious which parameter to choose (see e.g.~\cite{Koutserimpas:24}).  

In this work, we calculate electromagnetic wave scattering at a temporal interface for four different choices of explicitly time--varying Drude-Lorentz models, using the Laplace transform method developed in \cite{solís2021timevaryingmaterialspresencedispersion} to derive their temporal scattering coefficients.  In general we find the scattering amplitudes vary significantly between the models and are heavily dependent on the initial conditions of both field and medium.  However, these differences in the coefficients can be directly related to radiation from 'flashes' of current at the temporal interface, which arise from quantities that are discontinuous as a consequence of the different boundary conditions implied by the different model choices.  As part of these results we discuss some limitations with these models in the limit where the field--medium coupling (here the plasma frequency) is switched to zero.  A custom finite difference time domain (FDTD) algorithm was developed to verify the analytical predictions of each of the models.  Overall, we show that knowledge of the specific implementation of time--modulation is vitally important when modelling time-varying systems.

\section{Four Drude--Lorentz Models\label{sec:four-models}}

\begin{table*}
\begin{ruledtabular}
\begin{tabular}{ccc}
\\
 Model&Material Dynamics&Continuous Quantities\\ \hline
 \\
 1 & $\frac{d^2P(t)}{dt^2}+\gamma\frac{dP(t)}{dt}+\omega_0^2P(t)=\epsilon_0 \omega_p^2(t)E(t)$ & $\bf{P}$, $\frac{d\bf{P}}{dt}$, $\bf{E}$, $\frac{d\bf{E}}{dt}$, $\bf{D}$, $\frac{d\bf{D}}{dt}$\\
 \\
 2 & $\omega_p^2(t)\frac{d^2}{dt^2}\left(\frac{P(t)}{\omega_p^2(t)}\right)+\gamma\omega_p^2(t)\frac{d}{dt}\left(\frac{P(t)}{\omega_p^2(t)}\right)+\omega_0^2P(t)=\omega_p^2(t)\epsilon_0E(t)$ & $\frac{\bf{P}}{\omega_p^2(t)}$, $\frac{d}{dt}\left(\frac{\bf{P}}{\omega_p^2(t)}\right)$, $\bf{D}$, $\frac{d\bf{D}}{dt}$\\
 \\
 3 & $\omega_p^2(t)\frac{d}{dt}\left(\frac{1}{\omega_p^2(t)}\frac{dP}{dt}\right)+\gamma\frac{dP}{dt}+\omega_0^2\omega_p^2(t)\int_{t_0}^t \frac{1}{\omega_p^2(t)}\frac{dP}{dt}dt=\omega_p^2(t)\epsilon_0E(t)$ & $\bf{P}$, $\frac{1}{\omega_p^2(t)}\frac{d\bf{P}}{dt}$, $\bf{E}$, $\bf{D}$, $\frac{d\bf{D}}{dt}$\\
 \\
 4 & $\omega_p^2(t)\frac{d}{dt}\left(\frac{1}{\omega_p^2(t)}\frac{dP(t)}{dt}\right)+\gamma\frac{dP(t)}{dt}+\omega_0^2(t)P(t)=\epsilon_0\omega_p^2(t)E(t)$ & $\bf{P}$, $\frac{1}{\omega_p^2(t)}\frac{d\bf{P}}{dt}$, $\bf{E}$, $\bf{D}$, $\frac{d\bf{D}}{dt}$\\

\end{tabular}
\end{ruledtabular}
\caption{\label{tab:models}\textbf{Polarization dynamics in time--varying Drude--Lorentz models:}  Here we list the differential equations governing the dynamics of each of our four Drude--Lorentz models, along with the continuous quantities at a temporal discontinuity.  We find that all four models must conserve $\bf{D}$ and $\frac{d\bf{D}}{dt}$ across the boundary.}
\end{table*}

In this section we introduce our four models for a time--modulated dispersive material, which are summarised in Table \ref{tab:models}.  Without time--modulation, all models reduce to a standard Drude-Lorentz model where the frequency dependent relative permittivity $\epsilon_r$ is given by
\begin{equation}
    \epsilon_r=\epsilon_{\infty}+\frac{\omega_p^2}{\omega_0^2-\omega^2-{\rm i}\gamma\omega},
    \label{permittivity}
\end{equation}
with $\mu_{r}=1$.  In the following we assume $\epsilon_{\infty}=1$ and, to obtain tractable analytic results, take a lossless material $\gamma=0$ (see supplementary materials \S1 for the effects of $\gamma\neq0$).  In all cases we take the time dependence of our system to be induced by an abrupt, instantaneous change in plasma frequency from an initial value, $\omega_{p-}$, to a final value, $\omega_{p+}$.

\begin{figure}[H]
    \centering
    \includegraphics[width=0.8\linewidth]{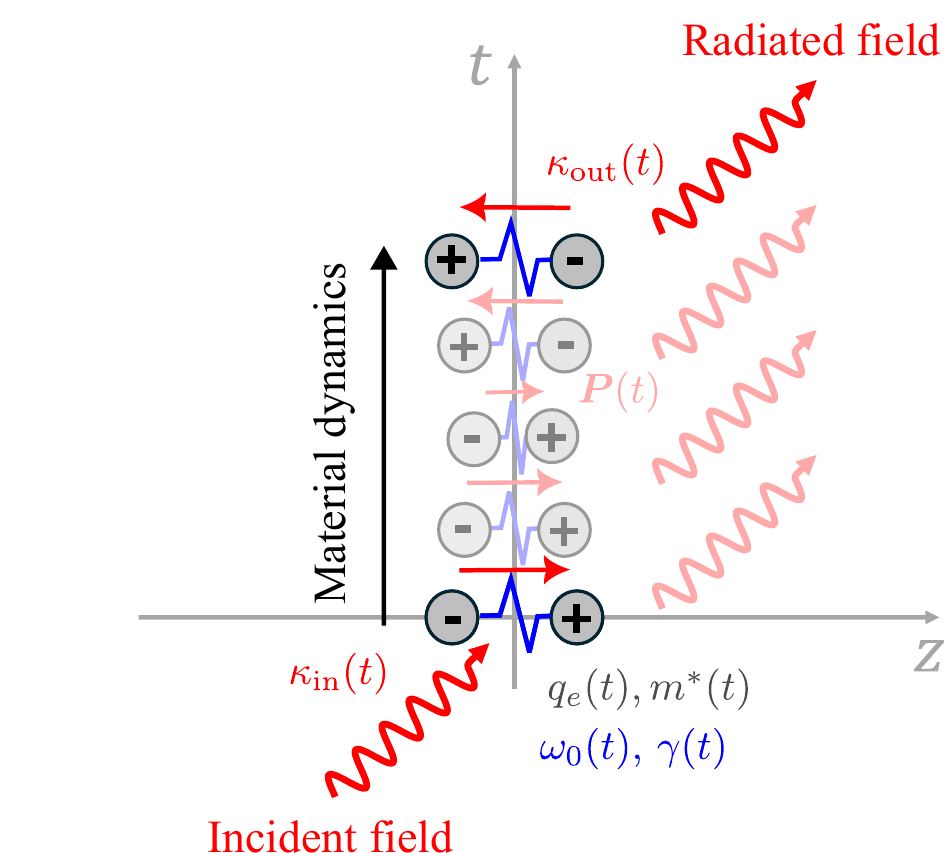}
    \caption{\textbf{Time--varying Drude-Lorentz models:} At each point $z$ within the medium an incident field couples to the harmonic motion of the polarization $\boldsymbol{P}$, with charge $q_e$, effective mass $m^{\star}$, resonant frequency $\omega_0$, and damping $\gamma$.  The coupling into the motion is proportional to $\kappa_{\rm in}$, and the coupling back out into the field is proportional to $\kappa_{\rm out}$.  For an abrupt change in the material parameters, we can achieve fixed initial and final permittivities, $\epsilon(\omega)$ through changing different combinations of these model parameters.  In model 1, the time modulation occurs through the `in--coupling', $\kappa_{\rm in}(t)$, in models 2 and 3 through the `out--coupling', $\kappa_{\rm out}(t)$, and in model 4, through the effective mass, $m^{\ast}(t)$.}
    \label{fig:coupling}
\end{figure}

Consider the case of an electromagnetic wave propagating in such a bulk medium.  Due to translational symmetry, momentum is conserved and thus the wave--vector $\boldsymbol{k}(\omega)=\frac{\omega}{c}\sqrt{\epsilon_r}\,\boldsymbol{e}_{z}$ must remain unchanged  We write this continuity condition as $k_-(\omega_-)=k_+(\omega_+)$, where the $\pm$ subscript indicates value before ($-$) and after ($+$) the switch.  Writing this momentum conservation condition out in terms of the permittivity (\ref{permittivity}) we thus have, 
\begin{equation}
    \frac{\omega_-}{c}\sqrt{1+\frac{\omega_{p-}^2}{\omega_0^2-\omega_-^2}} = \frac{\omega_+}{c}\sqrt{1+\frac{\omega_{p+}^2}{\omega_0^2-\omega_+^2}},
    \label{dispersion}
\end{equation}
where we've taken $\gamma=0$ as described above.  Once squared and the denominators cleared, this leads to a quartic polynomial in $\omega_+$, which can be solved straightforwardly, the roots being given by,
\begin{align}
\omega_+&=\pm\omega_{n}\nonumber\\
&=\pm\sqrt{\frac{K-(-1)^{n}\sqrt{K^2-\Delta}}{2(\omega_-^2-\omega_0^2)}},
    \label{quartic}
\end{align}
where we have defined,
\begin{equation}
    K = \omega_-^2(\omega_-^2+\omega_{p+}^2-\omega_{p-}^2)-\omega_0^2(\omega_0^2+\omega_{p+}^2),
\end{equation}
and
\begin{equation}
    \Delta=4\omega_0^2\omega_-^2(\omega_-^2-\omega_0^2)(\omega_-^2-\omega_0^2-\omega_{p-}^2).
\end{equation}
We denote the four roots (\ref{dispersion}) as $\pm\omega_1$ and $\pm\omega_2$, as defined in Eq. (\ref{quartic}).  If the plasma frequency varies suddenly in time then before and after this modulation, the wave will be in some superposition of these four frequencies.  We now proceed to show that the amplitudes in this superposition are heavily dependent on the model of the material dynamics used to describe the time modulation (see the schematic in Fig.~\ref{fig:coupling}).  We now summarize these four Drude--Lorentz models.

\subsection{Model 1: modulated in--coupling}

 We use the standard Drude--Lorentz model as our  starting point.  Here the polarization $\boldsymbol{P}(t)=\boldsymbol{e}_{x}P(t)$ obeys the equation of a damped simple harmonic oscillator with damping constant $\gamma$, resonant frequency $\omega_0$, and a coupling to the electric field proportional to the plasma frequency $\omega_{p}$,
\begin{equation}
    \frac{d^2P(t)}{dt^2}+\gamma\frac{dP(t)}{dt}+\omega_0^2P(t)=\epsilon_0 \omega_p^2E(t).
    \label{engheta}
\end{equation}
In Ref.~\cite{solís2021timevaryingmaterialspresencedispersion}, the time modulation of the material response is introduced through assuming a time-dependent plasma frequency $\omega_p(t)$, considering this to be equivalent to an abrupt change in the electron number density. The time-dependent plasma frequency is then defined as,
\begin{equation}
    \omega_p(t)=\sqrt{\frac{q_e^2 n(t)}{m^*\epsilon_0}},\label{eq:omega_p}
\end{equation}    
where $q_e$ is the electron charge, $m^*$ is the effective mass of the charge carriers, and $n(t)$ is the time-dependent electron number density given as $n(t)=n_-+(n_+-n_-)\Theta(t)$, where $\Theta(t)$ is the Heaviside step function.  The step function approximation holds for real systems, provided the duration of the change is sufficiently shorter than the period of the wave. Our first time--varying model for the polarization is therefore,
\begin{equation}
    \frac{d^2P(t)}{dt^2}+\gamma\frac{dP(t)}{dt}+\omega_0^2P(t)=\epsilon_0 \omega_p^2(t)E(t),
    \label{model 1}
\end{equation}
with Maxwell's equations unmodified.  Without singularities in the external current, Maxwell's equations imply that the $\boldsymbol{D}$ and $\boldsymbol{B}$ fields are continuous across a time interface~\cite{Galiffi_2022}, and from Eq. (\ref{model 1}) we observe---demanding that all time derivatives are finite---that the polarization and its time--derivative must also be continuous in this particular model. Through using the relation $\boldsymbol{D}=\epsilon_0\boldsymbol{E}+\boldsymbol{P}$, we thus also have continuity of $\boldsymbol{E}$ and $d\boldsymbol{E}/dt$, despite the carrier density $n(t)$ being discontinuous.

With reference to Fig.~\ref{fig:coupling}, we observe that this model can be understood as a simple harmonic oscillator with a time--varying in--coupling $\kappa_{\rm in}(t)$, of the field into the oscillatory motion, but a constant fixed out--coupling $\kappa_{\rm out}$ of the motion into radiation.

Note that, although we are writing the equations here in terms of electromagnetic parameters relevant for a condensed matter, or radio frequency metamaterial system, almost identical equations hold in acoustic systems, where constitutive relations can also be varied in time.  In that context, this first model is closely related to the time--modulated meta--atoms described in~\cite{Cho2020}, where a time--varying Drude--Lorentz response is achieved through modulating the in--coupling to the meta--atom (in-- and out--coupling here are respectively the volume of a speaker and microphone, coupled through a single--board computer).

\subsection{Model 2: modulated polarization density}

As the converse of model 1, we can envisage a situation where the in--coupling to the polarization field is fixed as $\kappa_{\rm in}$, while the out--coupling to the radiation $\kappa_{\rm out}(t)$, changes in time.  As we shall show, this choice has an effect on the frequency conversion when the permittivity is rapidly modulated, although the initial and final form of the frequency dependent permittivity is identical between the models.  Again, although we develop our formalism for electromagnetic waves, this situation can be implemented using the acoustic meta--atoms described in~\cite{Cho2020}, where the output volume of the speaker is varied in time.

We motivate this model through considering an individual charge of fixed effective mass $m^\ast$, at position $x(t)$, and undergoing harmonic motion, subject to the electric field, $\boldsymbol{E}=E\boldsymbol{e}_{x}$,
\begin{equation}
    \frac{d^2x(t)}{dt^2}+\gamma\frac{dx(t)}{dt}+\omega_0^2x(t)=\frac{q_e}{m^*}E(t).
    \label{electron_pos}
\end{equation}
We then introduce the polarisation $\boldsymbol{P}$ as the net dipole moment per unit volume inside the material, taking the density of carriers $n(t)$ to change in time such that,
\begin{equation}
    {\boldsymbol{P}}(t)=n(t)q_ex(t)\,\boldsymbol{e}_{x}.
    \label{mod2P}
\end{equation}
Combining the equation for the motion of the charge \eqref{electron_pos} with the definitions of both the polarization \eqref{mod2P} and the plasma frequency \eqref{eq:omega_p}, results in a differential equation in governing the evolution of the polarization,
\begin{multline}
    \omega_p^2(t)\frac{d^2}{dt^2}\left(\frac{P(t)}{\omega_p^2(t)}\right)+\gamma\omega_p^2(t)\frac{d}{dt}\left(\frac{P(t)}{\omega_p^2(t)}\right)\\ +\omega_0^2P(t)=\omega_p^2(t)\epsilon_0E(t),
    \label{model 2}
\end{multline}
which is identical to \eqref{model 1} when the plasma frequency is constant, reducing to the standard Drude--Lorentz model \eqref{engheta}.  However, examining \eqref{mod2P}, we can see that the continuous quantities are $\boldsymbol{P}/\omega_p^2$ and $d(\boldsymbol{P}/\omega_p^2)/dt$.  Given the continuity of $\boldsymbol{D}$ and $\boldsymbol{B}$ implied by Maxwell equations, a sudden switch in the plasma frequency results in a discontinuous $\boldsymbol{E}$ and $d\boldsymbol{E}/dt$ here.  As we shall see, this difference in the boundary conditions leads to very different temporal scattering coefficients. This version of the time-varying Drude-Lorentz model was used in Ref.~\cite{PhysRevA.111.033507}.

\subsection{Model 3: modulated current density}

Another version of model 2, where we vary the `out--coupling' in time, is written in terms of the current density $j=dP/dt$, rather than the polarization,
\begin{equation}
    \frac{dP(t)}{dt}=q_e\,n(t)\frac{dx(t)}{dt},
    \label{mod3P}
\end{equation}
where the current is given by the instantaneous carrier density $n(t)$, times the carrier charge and velocity.  This version of the time--varying Drude--Lorentz model was used in Ref.~\cite{PhysRevLett.130.203803}.  For static materials, Eq. \eqref{mod3P} is simply the time derivative of \eqref{mod2P}, again making the two models equivalent.  Using the expression for the current \eqref{mod3P} to write the equation of motion \eqref{electron_pos} in terms of the polarization field we now have an integro--differential equation for the polarization field,
\begin{multline}
    \omega_p^2(t)\frac{d}{dt}\left(\frac{1}{\omega_p^2(t)}\frac{dP}{dt}\right)+\gamma\frac{dP}{dt}+ \\ \omega_0^2\omega_p^2(t)\int_{t_0}^t \frac{1}{\omega_p^2(t)}\frac{dP}{dt}dt=\omega_p^2(t)\epsilon_0E(t).
    \label{model 3}
\end{multline}
where we take the boundary condition that the displacement $x(t)$ vanishes at time $t=t_0$.  Again, Eq. \eqref{model 3} reduces to the standard Drude--Lorentz model when $\omega_p$ is time independent.  However, note that the integral term makes the dynamics explicitly dependent on the history of the polarization field.  From an examination of \eqref{model 3}, both $P$ and $(1/\omega_p^2)\,dP/dt$ are continuous functions of time, implying that---for an abrupt change in plasma frequency---the electric field is continuous, while its time derivative is not.

\subsection{Model 4: modulated effective mass}

Finally we take the plasma frequency to vary through a change in the effective mass, $m^{\ast}$ of the charge carriers.  We modify the equation of motion \eqref{electron_pos}, remembering that the rate of change of momentum is given by $d(m^{\ast}\,dx/dt)/dt$.  Thus,
\begin{equation}
    \frac{d}{dt}\left(m^*(t)\frac{dx(t)}{dt}\right)+\gamma m^*(t)\frac{dx(t)}{dt}+\mathcal{K}x(t)=q_eE(t),
\end{equation}
where $\mathcal{K}(t)=m^{\ast}(t)\omega_0^2$ is the spring constant.

Following the same procedure described in the three previous cases and using \eqref{mod2P} to write the dynamics of the polarization field (taking constant value of carrier density $n$ in this model), this leads to,
\begin{multline}
    \omega_p^2(t)\frac{d}{dt}\left(\frac{1}{\omega_p^2(t)}\frac{dP(t)}{dt}\right)+\gamma\frac{dP(t)}{dt}+ \\ \omega_0^2 P(t)=\epsilon_0\omega_p^2(t)E(t).
    \label{model 4}
\end{multline}
Interestingly, when the resonant frequency $\omega_0$ equals zero, this model is equivalent to the modulated carrier density described in model 3. As the term proportional to $\omega_0$ in Eqns. \eqref{model 3} and \eqref{model 4} plays no role in determining the continuity conditions, the temporal boundary conditions here are identical to those in model 3.

\section{Scattering Coefficients}
\label{scattering_coefficients}

We now consider a plane wave of initial wave--vector $\boldsymbol{k}$ and frequency $\omega_{-}$ propagating in an infinite bulk medium where the polarization evolves according to one of the four models outlined in the previous section.  At $t=0$ the plasma frequency is rapidly switched from the value $\omega_{p-}$ to the value $\omega_{p+}$, causing the wave to scatter into the four different frequencies given by Eq. \eqref{quartic}.  To find the scattering amplitudes we use the Laplace transform method described in Ref.~\cite{solís2021timevaryingmaterialspresencedispersion}.  The full derivation of these amplitudes for each model are given in supplementary materials \S3.\\

We start from the vector wave equation obtained by eliminating the magnetic field from the two Maxwell equations involving curls, 
\begin{equation}
    \boldsymbol{\nabla} \times (\boldsymbol{\nabla} \times \boldsymbol{E}) = -\mu_0\epsilon_0\frac{\partial^2\boldsymbol{E}}{\partial t^2}-\mu_0\frac{\partial^2\boldsymbol{P}}{\partial t^2}.\label{eq:vector-wave}
\end{equation}
As the wave--vector is conserved in a homogeneous medium, we can make the replacement $\boldsymbol{\nabla}={\rm i}\boldsymbol{k}={\rm i}k\boldsymbol{e}_{z}$, assume the fixed polarization, $\boldsymbol{E}=E\boldsymbol{e}_{x}$ and $\boldsymbol{P}=P\boldsymbol{e}_{x}$, and thus reduce (\ref{eq:vector-wave}) to,
\begin{equation}
    \left(c^2k^2+\frac{d^2}{d t^2}\right)\epsilon_0 E=-\frac{d^2 P}{d t^2}.\label{eq:reduced-maxwell}
\end{equation}
Taking the Laplace transform of Eq. (\ref{eq:reduced-maxwell}) over all positive times $t>0$, we have,
\begin{multline}
    \left(c^2k^2+s^2\right)\epsilon_0 \tilde{E}(s) = \epsilon_0 s E(0^+)+\epsilon_0\frac{dE}{dt}(0^+) \\ -s^2\tilde{P}(s)+sP(0^+)+\frac{dP}{dt}(0^+).
    \label{laplace_field_eq}
\end{multline}
where $s$ is the Laplace transform variable, conjugate to time, and $0^{\pm}$ indicate infinitesimal positive/negative quantities.  Note that Eq. (\ref{laplace_field_eq}) represents the transform of the field and polarization \emph{after} the switch, where the medium is not changing any more.  The conversion between frequencies at the time boundary is thus entirely wrapped up in the connection between the electric field and polarization amplitudes and their time derivatives before and after the rapid change in plasma frequency (i.e. evaluated at the times $0^{+}$ and $0^{-}$). \\

To eliminate $\tilde{P}(s)$ from Eq. (\ref{laplace_field_eq}), we perform a second Laplace transform on one of the differential equations for the polarization dynamics, either Eqn. \eqref{model 1}, \eqref{model 2}, \eqref{model 3}, or \eqref{model 4}.  However, for times after the switch, all of these differential equations reduce to the standard Drude--Lorentz model for a static medium \eqref{engheta}.  Note that in model 3 this requires the boundary condition that the polarization and its time derivative vanish at $t=-\infty$, as discussed in supplementary materials \S3.  This results in an algebraic equation we can solve for the polarization, the solution of which is,
\begin{equation}
    \tilde{P}(s)=\frac{\epsilon_0\omega_{p+}^2\tilde{E}(s)+sP(0^+)+\frac{dP}{dt}(0^+)}{s^2+\omega_0^2} + \xi
    \label{laplace_p}
\end{equation}
where $\xi = 0$ for models 1, 2 and 4. Performing this Laplace transform over the integral term in model 3 results in $\xi = \frac{\omega_0^2}{s(s^2+\omega_0^2)}\left(P(0^+)-\frac{\omega_{p+}^2}{\omega_{p-}^2} P(0^-)\right)$. Substituting this into equation \eqref{laplace_field_eq} we find the relation between the Laplace amplitude of the electric field, and the values of the field and polarization immediately after the switch in the plasma frequency,
\begin{multline}
    \left[\left(c^2 k^2+s^2\right)(\omega_0^2+s^2)+s^2\omega_{p+}^2\right]\tilde{E}(s)\\=(s^2+\omega_0^2)\left[sE(0^+)+\frac{dE}{dt}(0^+)\right]\\+\frac{\omega_0^2}{\epsilon_0}\left[sP(0^+)+\frac{dP}{dt}(0^+)\right] - \frac{s^2(s^2+\omega_0^2)}{\epsilon_0}\xi.\label{eq:laplace-E}
\end{multline}
Before the abrupt change in the plasma frequency at $t=0$, we take the electric field as a unit amplitude travelling wave, and find the polarization as the inhomogeneous solution to Eq. (\ref{model 1}) with a constant value of the plasma frequency $\omega_p=\omega_{p-}$.  These expressions are,
\begin{align}
    E(t<0)&=\cos(kz-\omega_-t)\nonumber\\[5pt]
    P(t<0)&=\frac{\epsilon_0\omega_{p-}^2}{\omega_0^2-\omega_-^2}\cos(kz-\omega_-t).\label{eq:E-P-before}
\end{align}
Taking the time $t=0^-$, just before the discontinuous jump in the plasma frequency, Eq. (\ref{eq:E-P-before}) shows that $E(0^-)=\cos(kz)$; $\left.dE/dt\right|_{0^-}=\omega_-\sin(kz)$; $P(0^-)=\epsilon_0\omega_{p-}^2 \cos(kz)/(\omega_0^2-\omega_-^2)$; and $\left.dP/dt\right|_{0^-}=\epsilon_0\omega_-\omega_{p-}^2 \sin(kz)/(\omega_0^2-\omega_-^2)$.  Table~\ref{tab:models} then tells us how to relate these values to those after the jump in plasma frequency.  These values are given in Table~\ref{tab:quantities}.
\begin{table*}
\begin{ruledtabular}
\begin{tabular}{ccccc}
\\
 Quantity&Model 1&Model 2&Model 3&Model 4\\ \hline
 $E(0^+)=$   &   $\cos(kz)$ & $\left(1+\frac{\omega_{p-}^2-\omega_{p+}^2}{\omega_0^2-\omega_-^2}\right)\cos(kz)$ & $\cos(kz)$ & $\cos(kz)$\\
 $\frac{dE}{dt}(0^+)=$   &   $\omega_- \sin(kz)$ & $\left(1+\frac{\omega_{p-}^2-\omega_{p+}^2}{\omega_0^2-\omega_-^2}\right)\omega_-\sin(kz)$ & $\left(1+\frac{\omega_{p-}^2-\omega_{p+}^2}{\omega_0^2-\omega_-^2}\right)\omega_-\sin(kz)$&$\left(1+\frac{\omega_{p-}^2-\omega_{p+}^2}{\omega_0^2-\omega_-^2}\right)\omega_-\sin(kz)$\\
 $P(0^+)=$   &   $\frac{\epsilon_0\omega_{p-}^2}{\omega_0^2-\omega_-^2}\cos(kz)$ & $\frac{\epsilon_0\omega_{p+}^2}{\omega_0^2-\omega_-^2}\cos(kz)$&$\frac{\epsilon_0\omega_{p-}^2}{\omega_0^2-\omega_-^2}\cos(kz)$&$\frac{\epsilon_0\omega_{p-}^2}{\omega_0^2-\omega_-^2}\cos(kz)$\\
 $\frac{dP}{dt}(0^+)=$   &  $\frac{\epsilon_0\omega_-\omega_{p-}^2}{\omega_0^2-\omega_-^2}\sin(kz)$ &  $\frac{\epsilon_0\omega_-\omega_{p+}^2}{\omega_0^2-\omega_-^2}\sin(kz)$ & $\frac{\epsilon_0\omega_-\omega_{p+}^2}{\omega_0^2-\omega_-^2}\sin(kz)$ & $\frac{\epsilon_0\omega_-\omega_{p+}^2}{\omega_0^2-\omega_-^2}\sin(kz)$
 \\
\end{tabular}
\end{ruledtabular}
\caption{\textbf{Field amplitudes immediately after the time boundary:} Electric field, polarization, and their time derivatives immediately after an abrupt change in the plasma frequency, from $\omega_{p-}$ to $\omega_{p+}$ at $t=0$.  These values determine the Laplace amplitude of the electric field via Eq. (\ref{eq:laplace-E}).\label{tab:quantities}}
\end{table*}

\begin{figure}[h!]
\includegraphics[width=\linewidth]{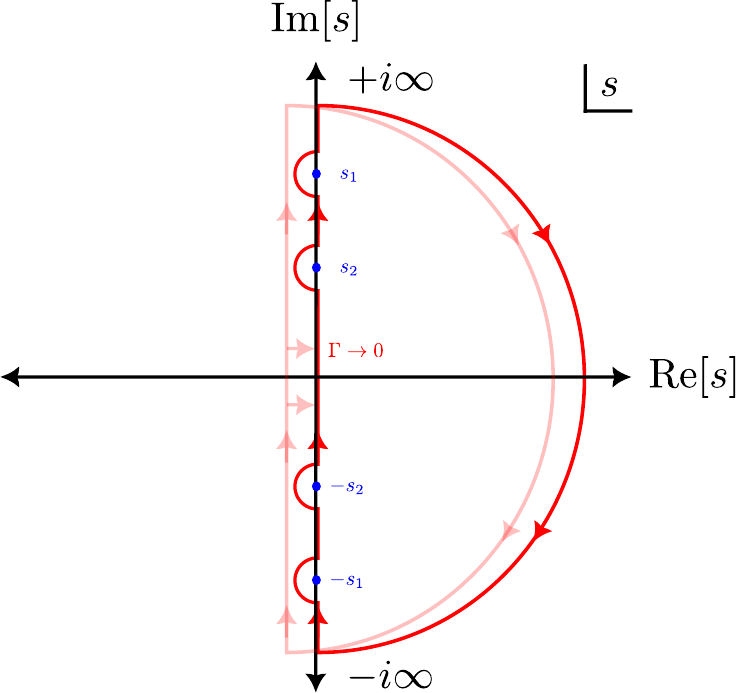}
\caption{\textbf{Inverse Laplace Transform Contour:} To perform the inverse Laplace transform in Eq. \eqref{eq:inverse_laplace_result} we complete the integration contour with a semi circle in the right hand plane.  We take $\Gamma<0$, so that the four poles are captured within the contour (contributing $-2\pi {\rm i}$ times each residue), equivalent to imposing a decay onto the four waves propagating within the material after the switch.  The $\Gamma\to0$ limit is then taken, meaning that the contour must pass clockwise around each pole as we integrate along the imaginary $s$ axis.\label{fig:laplace_contour}}
\end{figure}

After substituting the values listed in Table \ref{tab:quantities} into Eq. (\ref{eq:laplace-E}), we have completely determined the electric field in all of the models, before and after the switch in plasma frequency.  Performing the inverse Laplace transform using the expression for $\tilde{E}(s)$ given in \eqref{eq:laplace-E} we have
\begin{align}
    E(t)&=\int_{-{\rm i}\infty+\Gamma}^{{\rm i}\infty+\Gamma}\frac{ds}{2\pi {\rm i}}\,\tilde{E}(s) e^{st}\nonumber\\[5pt]
    &=\frac{F(s_1)\,e^{s_1 t}-F(-s1)\,e^{-s_1t}}{2s_1(s_2^2-s_1^2)}\nonumber\\[5pt]
    &\hspace{3cm}-\frac{F(s_2)e^{s_2t}-F(-s_2)e^{-s_2t}}{2s_2(s_2^2-s_1^2)}\label{eq:inverse_laplace_result}
\end{align}
where $\Gamma$ is a number we choose to ensure the convergence of the integral, we have defined $F(s)=(s^2+\omega_0^2)[sE(0^+)+dE/dt(0^+)]+\epsilon_0^{-1}\omega_0^2[sP(0^+)+dP/dt(0^+)]-\epsilon_0^{-1}s^2(s^2-\omega_0^2)\xi$, and $\pm s_1$ and $\pm s_2$ are the four imaginary roots of the polynomial $(c^2k^2+s^2)(\omega_0^2+s^2)+s^2\omega_{p+}^2=0$.  In Fig. \ref{fig:laplace_contour} we show the contour used to evaluate the integral in Eq. \eqref{eq:inverse_laplace_result}.  The quantity $F$ within each of the terms within Eq. (\ref{eq:inverse_laplace_result}) depends on the initial values of the electric field and polarization, which---as we have already discussed---are model dependent.
%
%

We can see that Eq. \eqref{eq:inverse_laplace_result} is a solution to the wave equation of the form,
\begin{equation}
    E(t)=E_1^++E_1^-+E_2^++E_2^-.
\end{equation}
with each wave evolving in time with the frequency $\pm \omega_{1,2}$ defined in Eq. \eqref{quartic}.  The amplitudes of each of these waves are the four temporal scattering coefficients.\\

\begin{figure}[H]
    \centering
    \includegraphics[width=\linewidth]{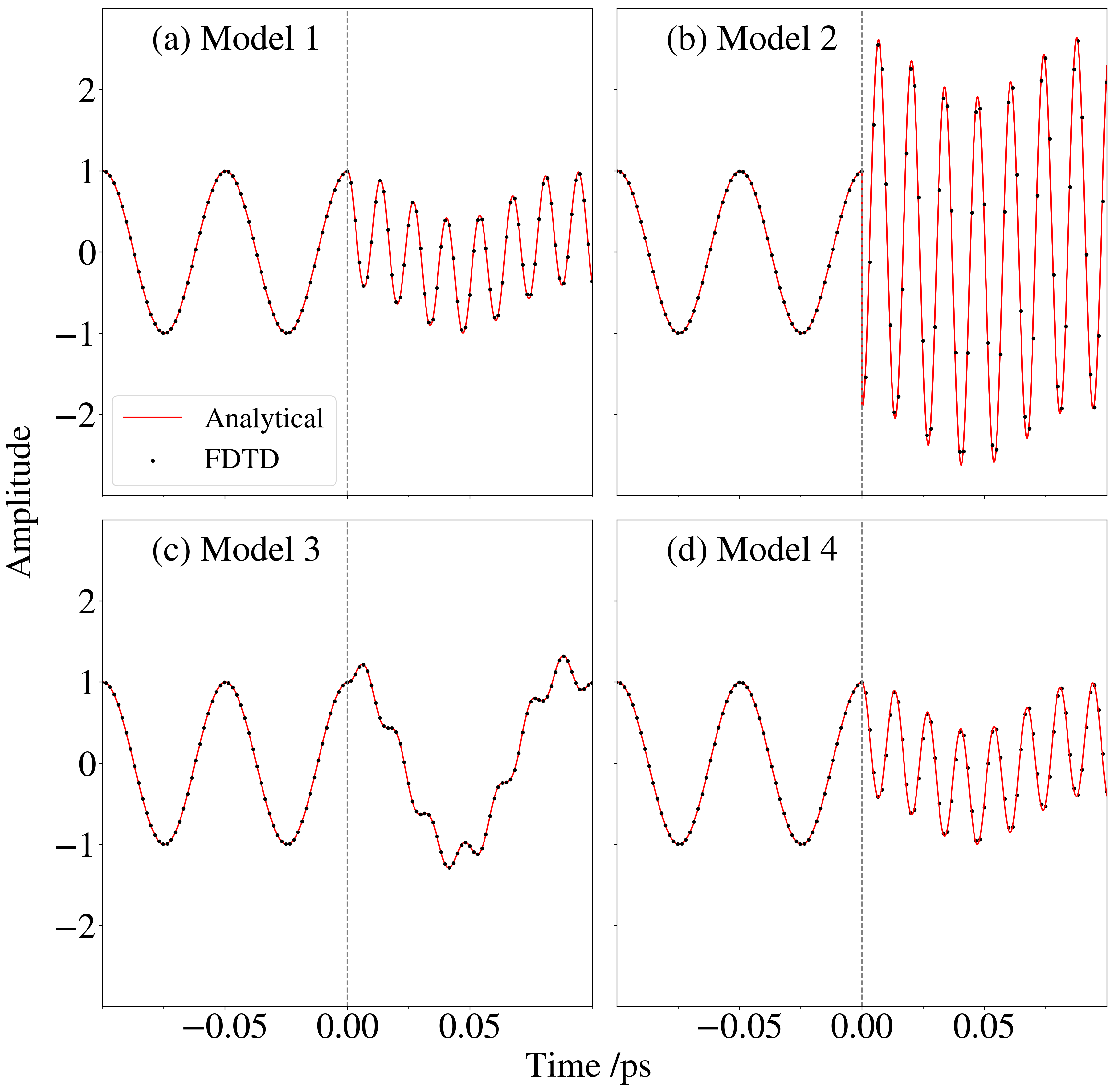}
    \caption{\textbf{Evolution of the electric field in different time--varying Drude--Lorentz models:}  The electric field evaluated at a fixed point within the material for each of the four models described in Sec. \ref{sec:four-models}, with $\omega_-=2\times10^{13}(2\pi)$, $\omega_0=2\omega_-$, $\omega_{p-}=0.5\omega_-$, $\omega_{p+}=3\omega_-$ and $\gamma=0$. The dashed line represents the temporal boundary at $t=0$. Before the interface, the wave is taken as identical in all 4 models, after the interface the wave splits into 2 forward and 2 backward spatially propagating waves with 2 unique frequencies. The black dots represent the FDTD simulation whilst the red line represents the analytically derived solutions from the Laplace transform method.}
    \label{fig:Ey}
\end{figure}

To verify these solutions, we used a custom finite difference time domain (FDTD) algorithm for each of the Drude-Lorentz models.  To examine the scattering coefficients, we injected a single frequency wave with unit amplitude into a material with plasma frequency $\omega_{p-}$. After a time $t_0$, the plasma frequency of the entire system was instantaneously switched to $\omega_{p+}$, and the amplitudes and frequencies of the four scattered waves were compared to those of the theory. Figure \ref{fig:Ey} shows the electric field output of the FDTD simulation in the time domain over-plotted with the predicted electric field solution according to the Laplace transform theory for each model for a particular set of system parameters. We see that our analytical solutions are in direct agreement with the simulation for all four models.
%
%
%
%
\section{Comparison between models: emission at the interface}

Whilst the full electric field after the jump in plasma frequency is given in Eq. \eqref{eq:inverse_laplace_result}, we now compare the scattering coefficients of all four models relative to model 1, which was that used in Ref.~\cite{solís2021timevaryingmaterialspresencedispersion}.  Here, we refer to the forward and backward scattering coefficients representing the waves with frequencies $\pm \omega_1$ and $\pm \omega_2$ as $A_{1,2}^\pm$. We denote the scattering coefficients for a particular model with a Roman numeral (I, II, III, or IV). Firstly the waves travelling with $\pm\omega_1$ are given by
\begin{align}
        A_{1,{\rm II}}^\pm &= A_{1,{\rm I}}^\pm + \frac{\left(\omega_1^2\pm\omega_-\omega_1\right)\left(\omega_{p+}^2-\omega_{p-}^2\right)}{2\left(\omega_-^2-\omega_0^2\right)\left(\omega_1^2-\omega_2^2\right)},\label{eq:A2p}\\[5pt]
        A_{1,{\rm III}}^\pm &= A_{1,{\rm I}}^\pm + \frac{\left(\omega_0^2\pm\omega_-\omega_1\right)\left(\omega_{p+}^2-\omega_{p-}^2\right)}{2\left(\omega_-^2-\omega_0^2\right)\left(\omega_1^2-\omega_2^2\right)},\label{eq:A3p}\\[5pt]
        A_{1,{\rm IV}}^\pm &= A_{1,{\rm I}}^\pm \pm \frac{\omega_-\omega_1\left(\omega_{p+}^2-\omega_{p-}^2\right)}{2\left(\omega_-^2-\omega_0^2\right)\left(\omega_1^2-\omega_2^2\right)}.\label{eq:A4p}
\end{align}
Similarly, the scattering coefficients for the waves travelling with $\pm\omega_2$ equal
\begin{align}
        A_{2,{\rm II}}^\pm &= A_{2, {\rm I}}^\pm - \frac{\left(\omega_2^2\pm\omega_-\omega_2\right)\left(\omega_{p+}^2-\omega_{p-}^2\right)}{2\left(\omega_-^2-\omega_0^2\right)\left(\omega_1^2-\omega_2^2\right)},\\[5pt]
        A_{2,{\rm III}}^\pm &= A_{2,{\rm I}}^\pm - \frac{\left(\omega_0^2\pm\omega_-\omega_2\right)\left(\omega_{p+}^2-\omega_{p-}^2\right)}{2\left(\omega_-^2-\omega_0^2\right)\left(\omega_1^2-\omega_2^2\right)},\\[5pt]
        A_{2,{\rm IV}}^\pm &= A_{2,{\rm I}}^\pm \mp \frac{\omega_-\omega_2\left(\omega_{p+}^2-\omega_{p-}^2\right)}{2\left(\omega_-^2-\omega_0^2\right)\left(\omega_1^2-\omega_2^2\right)}.\label{eq:A2final}
\end{align}
The dependence of these coefficients on the difference in plasma frequency is explored in Figs. \ref{fig:no_res}--\ref{fig:res_2}.  As expected, the difference between the scattering coefficients is proportional to the difference in the initial and final squared plasma frequencies, $\omega_{p+}^2-\omega_{p-}^2$, vanishing when there is no time modulation of the material.  Similarly, the difference between the models appears to become uniformly large when the two final frequencies become degenerate $\omega_{2}^2\to\omega_1^2$.  This is the usual situation when two roots of a dispersion relation coalesce, indicating that the contour integral shown in Fig. \ref{fig:laplace_contour} contains two double poles rather than four single poles, meaning that the residue in Eq. \eqref{eq:inverse_laplace_result} must be extracted in terms of the value of the derivative $dF(s)/ds$, rather than the value of $F(s)$ at the pole.  There is also a divergence in the amplitudes when $\omega_{-}\to\omega_{0}$, which is when the wave is resonant with the polarization dynamics.  Therefore the only significant difference in the models arises from the remaining factors of e.g. $(\omega_{1}^2\pm\omega_-\omega_1)$, $(\omega_0^2\pm\omega_-\omega_1)$, and $\omega_-\omega_1$ in Eqns. (\ref{eq:A2p}--\ref{eq:A4p}).  In the next two sections we discuss the origin of these differences in the scattering amplitudes in more detail.

\subsection{Drude models ($\omega_0=0$):}

\begin{figure*}
    \centering
    \includegraphics[width=\linewidth]{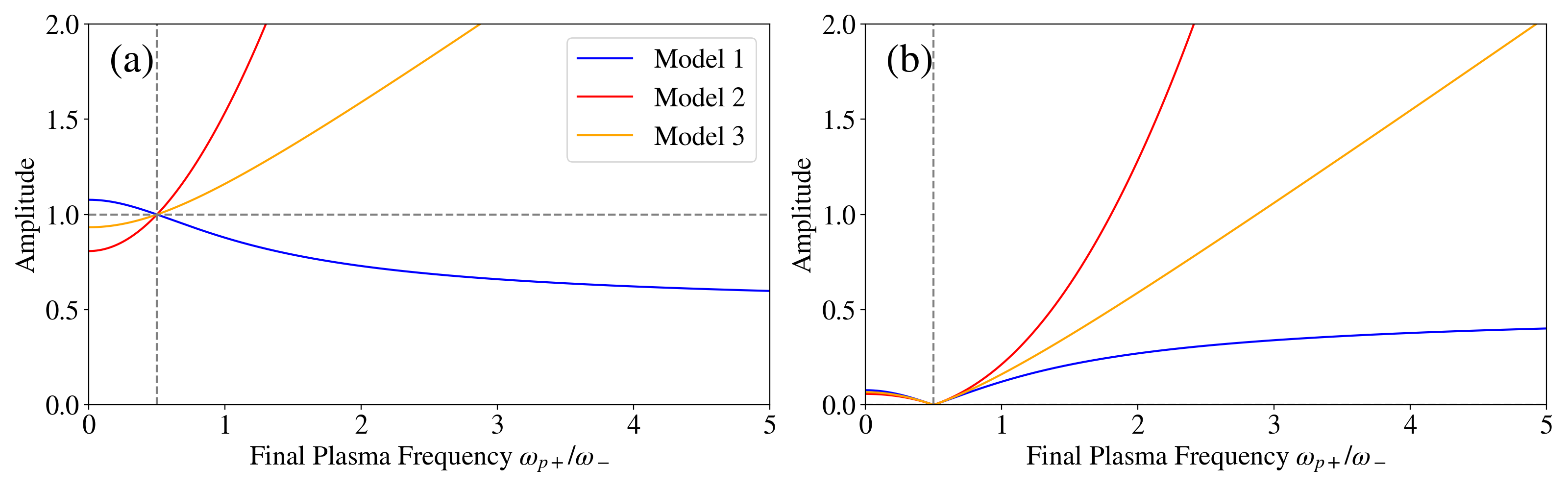}
    \caption{\textbf{Forward and Backward Amplitudes $A_1^{\pm}$ for different time--varying Drude models ($\omega_{p-}=0.5\omega_-$, $\omega_0=0$):} (a) forward $A_1^+$ and (b) backward $A_1^-$ scattering coefficients as a function of the final plasma frequency.  All frequencies are given in terms of the initial frequency $\omega_-$, in a system with initial conditions given by $\omega_{p-}=0.5\omega_-$, $\omega_0=0$. The coloured lines represent the scattering coefficients for each of the different models. For this case with $\omega_0=0$, model 4 is identical to model 3 so we only show the results for model 3. The vertical dashed lines represent no switching ($\omega_{p-}=\omega_{p+}$) and the horizontal dashed lines show the amplitude of the initial wave. We see a clear trend in the order of the models for $\omega_{p+}>\omega_{p-}$ with the opposite being true for $\omega_{p+}<\omega_{p-}$. Most importantly, we see the predicted reciprocal, quadratic, and linear behaviours present in models 1,2 and 3 transformed by the initial conditions.}
    \label{fig:no_res}
\end{figure*}

\begin{figure*}
    \centering
    \includegraphics[width=\linewidth]{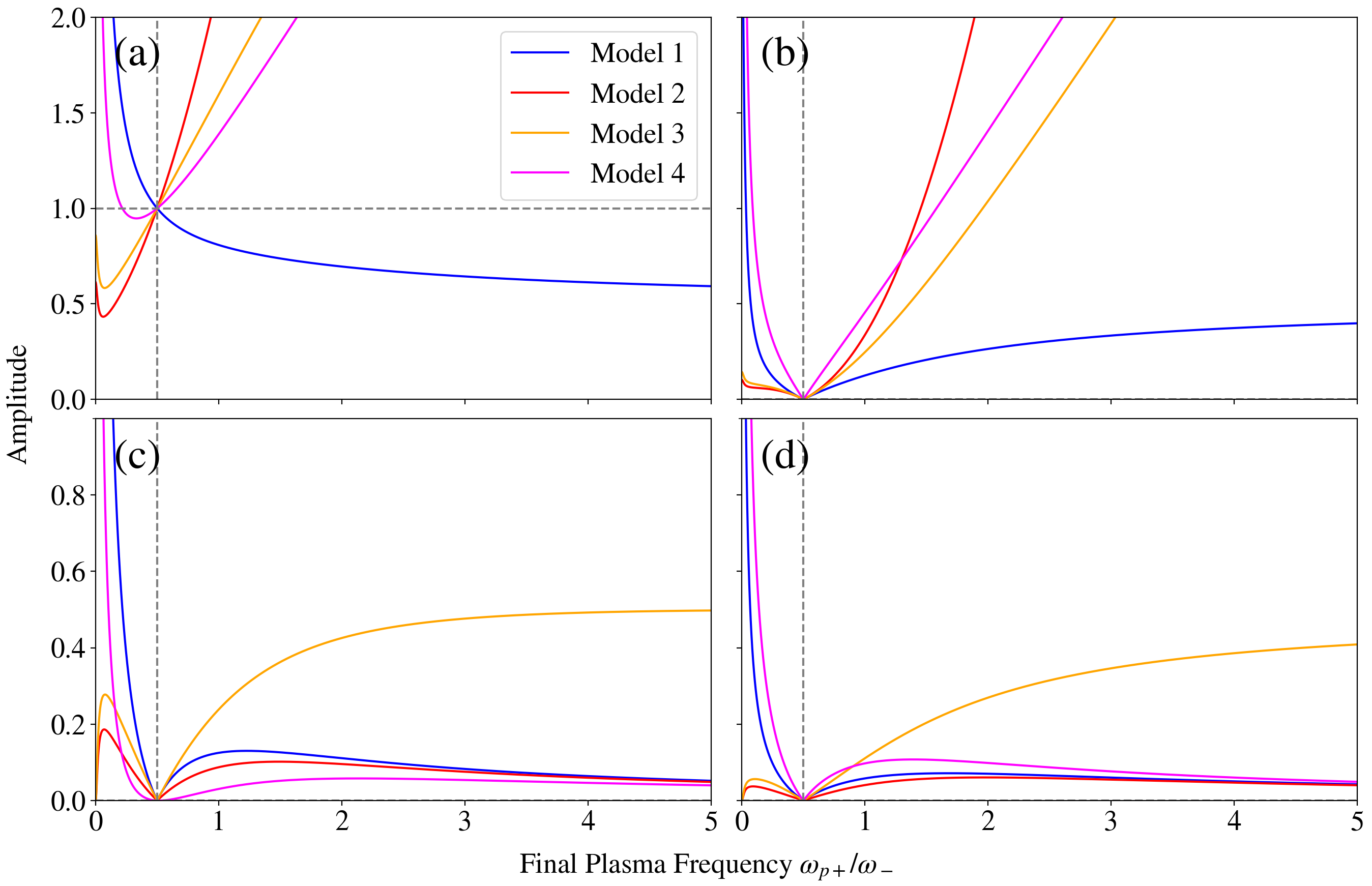}
    \caption{\textbf{Forward and backward scattering coefficients, $A_{1}^{\pm}$ and $A_{2}^{\pm}$ for different time--varying Drude--Lorentz models ($\omega_{p-}=0.5\omega_-$ and $\omega_0=0.7\omega_-$):} As for Fig. \ref{fig:no_res}, with panels (a), (b), (c) and (d) showing the scattering coefficients $A_1^+$, $A_1^-$, $A_2^+$, $A_2^-$.}
    \label{fig:res_07}
\end{figure*}

\begin{figure*}
    \centering
    \includegraphics[width=\linewidth]{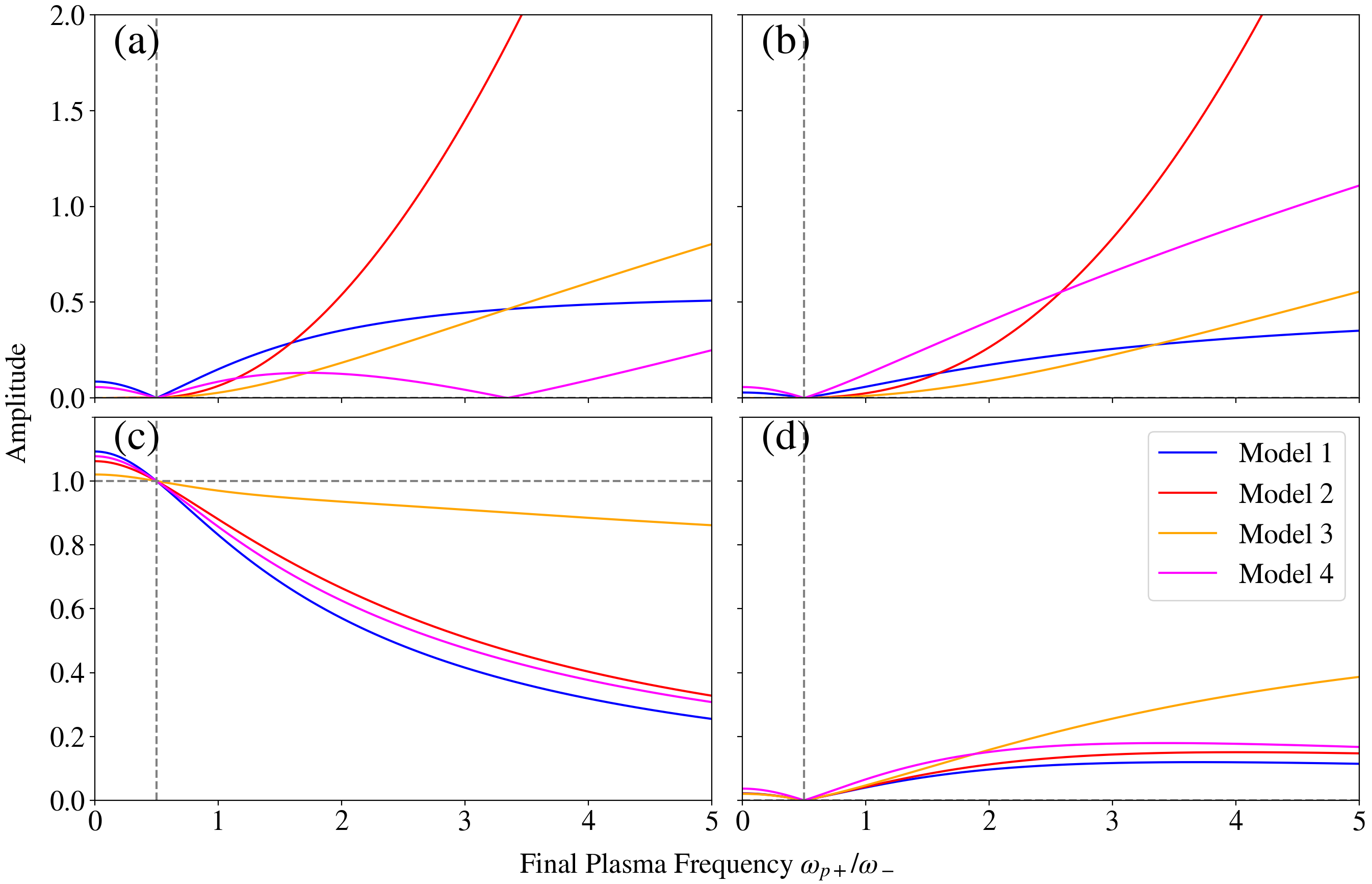}
    \caption{\textbf{Forward and backward scattering coefficients $A_{1}^{\pm}$ and $A_{2}^{\pm}$ for different time--varying Drude--Lorentz models ($\omega_{p-}=0.5\omega_-$ and $\omega_0=2\omega_-$):}  Except for the modified model parameters, the axes and meaning of the curves are the same as for Figs. \ref{fig:no_res} and \ref{fig:res_07}.}
    \label{fig:res_2}
\end{figure*}

For simplicity, we start by examining the case where the Drude--Lorentz model reduces to the Drude model, $\omega_0=0$.  In this case the four scattering coefficients described above reduce to two, with the wave frequencies given by $\omega_1=\pm(\omega_-^2+\omega_{p+}^2-\omega_{p-}^2)^{1/2}$, and $\omega_2=0$.  Note that models 3 and 4 are identical in this case, so we exclude model 4. In this case---from Eqns. (\ref{eq:A2p}--\ref{eq:A4p})---the $A_1^+$ scattering coefficients simplify to
\begin{align}
        A_{1,{\rm I}}^+ &= \frac{1}{2} + \frac{\omega_-}{2\sqrt{\omega_-^2+\omega_{p+}^2-\omega_{p-}^2}},\nonumber\\[5pt]
        A_{1,{\rm II}}^+ &= \frac{1}{2} + \frac{\omega_-}{2\sqrt{\omega_-^2+\omega_{p+}^2-\omega_{p-}^2}}\nonumber \\ &+\frac{\omega_{p+}^2-\omega_{p-}^2}{2\omega_-\sqrt{\omega_-^2+\omega_{p+}^2-\omega_{p-}^2}} + \frac{\omega_{p+}^2-\omega_{p-}^2}{2\omega_-^2},\nonumber\\[5pt]
        A_{1,{\rm III}}^+ &= \frac{1}{2} + \frac{\omega_-}{2\sqrt{\omega_-^2+\omega_{p+}^2-\omega_{p-}^2}}\nonumber\\ &+\frac{\omega_{p+}^2-\omega_{p-}^2}{2\omega_-\sqrt{\omega_-^2+\omega_{p+}^2-\omega_{p-}^2}}.\label{no_res}
\end{align}
To illustrate the origin of the additional terms in the results of the previous section (\ref{eq:A2p}--\ref{eq:A4p}), we only consider the above positive frequency amplitudes. An identical analysis can be performed for their negative frequency counterparts, but we do not include this here.  Note that in the limit of a large jump in the value of the plasma frequency, $\omega_{p+}\to\infty$, we have $A^{+}_{1,{\rm I}}\to1/2$, $A^{+}_{1,{\rm II}}\to\omega_{p+}^2/2\omega_-^2$, and $A^+_{1,{\rm III}}\to |\omega_{p+}|/2\omega_-$.  This difference between the models is clear in Fig.~\ref{fig:no_res}, where the respective choices lead to a constant, quadratic, or linear scaling with $\omega_{p+}$ in the large contrast limit.

To give a physical understanding of the additional terms in the models given in Eq. (\ref{no_res}), along with their scaling with respect to the final plasma frequency, we consider the time derivative of the current in each model of the polarization dynamics, this being responsible for radiation.  In each case $dj/dt=d^2P/dt^2$ equals,

\begin{align}
        \left[\frac{d^2P}{dt^2}\right]_{{\rm I}}&=\omega_p^2(t)\epsilon_0E(t),\nonumber\\[5pt]
        \left[\frac{d^2P}{dt^2}\right]_{\rm II}&=\omega_p^2(t)\epsilon_0E(t) + 2\frac{d\omega_p^2(t)}{dt}\epsilon_0\int_{t_0}^tE(t)dt \nonumber\\ &+\frac{d^2\omega_p^2(t)}{dt^2}\epsilon_0 \iint_{t_0}^t E(t)dt,\nonumber\\[5pt]
        \left[\frac{d^2P}{dt^2}\right]_{\rm III}&=\omega_p^2(t)\epsilon_0E(t) + \frac{d\omega_p^2(t)}{dt}\epsilon_0\int_{t_0}^tE(t)dt.\label{eq:currents}
\end{align}

As discussed above, here the plasma frequency is taken to change abruptly at $t=0$.  We can thus see that, whereas model 1 corresponds to a finite, albeit discontinuous current density, model 2 contains both a first and second order time derivative of the squared plasma frequency, corresponding to a delta function and its time derivative.  Model 3 also contains only a first order derivative of the squared plasma frequency, and thus also a delta function.  Physically these delta functions and their derivatives represent `flashes' of current that occur as the plasma frequency is abruptly changed.  Of course, in reality any change in the plasma frequency will be over a non--zero time interval $t_{\rm rise}$, and these derivatives will be replaced by terms proportional to inverse powers of this rise time, $t_{\rm rise}^0$, $t_{\rm rise}^{-1}$, and $t_{\rm rise}^{-2}$.  However, we can see that when there is a large abrupt change in the plasma frequency, we should expect larger temporal scattering coefficients for model 2, followed by model 3, and then finally model 1.  This is exactly what we observe in both panels of Fig. \ref{fig:no_res}.

We now show that the radiation from the additional temporal interface currents identified in Eq. \eqref{eq:currents} does indeed correspond to the additional terms within the scattering amplitudes of the various models listed in Eq. \eqref{no_res}. Using the Green function for the simple harmonic oscillator we find the contribution to the electric field (see supplementary materials \S4), for the interface currents in models 2 and 3 arising from the additional terms dependent on derivatives of the plasma frequency,

\begin{widetext}
\begin{align}
        E_{{\rm flash}, {\rm II}}(t)&=\cos(kz-\omega_1t)\Theta(t)\left(\frac{\omega_{p+}^2-\omega_{p-}^2}{2\omega_-^2}+\frac{\omega_{p+}^2-\omega_{p-}^2}{2\omega_-\sqrt{\omega_-^2+\omega_{p+}^2-\omega_{p-}^2}}\right),\\
        E_{{\rm flash}, {\rm III}}(t)&=\cos(kz-\omega_1t)\Theta(t)\left(\frac{\omega_{p+}^2-\omega_{p-}^2}{2\omega_-\sqrt{\omega_-^2+\omega_{p+}^2-\omega_{p-}^2}}\right).
    \label{current_amps}
\end{align}
\end{widetext}
As anticipated, these additional terms are identical to those found in the scattering coefficients \eqref{no_res}, when compared with model 1.  We can thus interpret the different temporal scattering in the choices of time--varying dispersive model as being due to additional currents at the temporal interface, currents that arise from their dependence on time derivatives of the plasma frequency.
%
%
%
%
%
%
\subsection{Drude--Lorentz models ($\omega_0\neq0$):}

We now extend the results of the previous section to the Drude--Lorentz model, where $\omega_0\neq0$.  Figs. \ref{fig:res_07} and \ref{fig:res_2} are examples plots of the scattering amplitudes in systems where  $\omega_0<\omega_-$  and $\omega_0>\omega_-$ respectively, where we note that the scattering amplitudes have a very different dependence on $\omega_{p+}$.

Unlike the Drude model ($\omega_0=0$), the scattering amplitudes are more complex in this case.  To simplify the discussion, we consider the $A_1^+$ scattering amplitudes in the limit of $\omega_{p+}\rightarrow\infty$. In this limit we have $\omega_1\rightarrow\omega_{p+}$ and $\omega_2\rightarrow0$, so that the amplitudes become
\begin{align}
        A_{1,{\rm I}}^+ &= \frac{1}{2},\nonumber\\[5pt]
        A_{1,{\rm II}}^+ &= \frac{1}{2} + \frac{\omega_-\omega_{p+}}{2\left(\omega_-^2-\omega_0^2\right)}+\frac{\omega_{p+}^2}{2\left(\omega_-^2-\omega_0^2\right)},\nonumber\\[5pt]
        A_{1,{\rm III}}^+ &= \frac{\omega_-^2}{2\left(\omega_-^2-\omega_0^2\right)}+\frac{\omega_-\omega_{p+}}{2\left(\omega_-^2-\omega_0^2\right)},\nonumber\\[5pt]
        A_{1,{\rm IV}}^+ &= \frac{1}{2}+\frac{\omega_-\omega_{p+}}{2\left(\omega_-^2-\omega_0^2\right)}.
        \label{eq:drude-lorentz-scattering}
\end{align}

By following the same reasoning as the above analysis of the Drude model, we can again show that the additional terms in the scattering amplitudes originate from an additional `flash' of current at the boundary, due to the time derivative of the plasma frequency.  Through expanding out the model equations \eqref{model 1}, \eqref{model 2}, \eqref{model 3}, \eqref{model 4} we can find the time derivative of the electric current, $\frac{d^2P}{dt^2}$, then we again use the simple harmonic oscillator Green function to find the electric field radiated from these forms of the current. The full derivation is detailed in supplementary materials \S5 where we prove these 'flash' terms solve to
\begin{align}
    E_{{\rm flash}, \rm II}(t) &= \left(\frac{\omega_-\omega_{p+}}{2(\omega_-^2-\omega_0^2)}+\frac{\omega_{p+}^2}{2(\omega_-^2-\omega_0^2)}\right)\cos(kz - \omega_{p+}t),\nonumber\\
    E_{{\rm flash}, \rm III}(t) &= \frac{\omega_-\omega_{p+}}{2(\omega_-^2-\omega_0^2)}\cos(kz-\omega_{p+}t),\nonumber\\
    E_{{\rm flash}, \rm IV}(t) &= \frac{\omega_{p+}\omega_-}{2(\omega_-^2-\omega_0^2)}\cos(kz-\omega_{p+}t).
\end{align}

We see that these terms are exactly as predicted in \eqref{eq:drude-lorentz-scattering}. As model 3 follows a different continuous equation, this gives rise to a different constant term at the start. We thus conclude that the differences in the efficiency of frequency conversion in dispersive models of time--varying media, arise from additional currents at any given temporal interface.  This difference in the electrical current is a consequence of the ambiguity in what is actually varying when we say that e.g. the plasma frequency is time dependent.  Is it the carrier density, the mass, or simply the influence of the electric field on the polarization dynamics?

%
%
\section{Discussions and Conclusions}

In this work, we have examined four commonly used time-varying Drude-Lorentz models. Using the Laplace transform method, we have determined the temporal scattering coefficients for a step function-like change in plasma frequency in an undamped system for all four models, finding that the scattering coefficients of the models are significantly different, particularly as the change in plasma frequency grows larger.  We have shown that the behaviours of the models are highly dependent on the initial conditions ($\omega_0$, $\omega_-$, $\omega_{p-}$) and that they each have a unique dependence on the change in plasma frequency given as reciprocal, linear or quadratic growth with the ratio $\omega_{{\rm p}+}/\omega_{-}$, depending on the model.  We have proven that these differences occur due to additional terms in the current derivatives, attributed to sudden flashes of current at the temporal interface. These results were verified using a custom FDTD algorithm, which we have additionally used to show some key limitations of these models in supplementary material \S2, whereby all models predict some form of unphysical behaviour when the exact radiative damping is unknown. \\

There is potential to experimentally test the applicability of these models to time varying media in two regimes. In optics, it is possible to change the effective plasma frequency of e.g. graphene using an intense laser pulse. Sufficiently high intensity laser pulses would excite electrons from the valence band into the conduction band, therefore modifying the electron number density akin to models two  and three.  Meanwhile in ITO, \cite{Jaffray:22, Bohn2021, LustigSegalSahaBordoChowdhurySharabiFleischerBoltassevaCohenShalaevSegev+2023+2221+2230}, the dominant effect of an incident pulse is to heat up the electrons in the conduction band, thus modifying the effective mass of the band, which is something like model four.  A more flexible experimental test of the above described model-dependent temporal scattering coefficients can be performed in the acoustic regime, making use of acoustic meta-atoms as described in e.g.~\cite{Cho2020}. As already shown there, the Drude--Lorentz model can be implemented via electronic feedback, and the `in--coupling' and `out--coupling' between the `atom' and the field (sketched in Fig. \ref{fig:coupling}) simply correspond to the sensitivity of the microphone and speaker, respectively. \\

Overall, we find that---in the regime of rapid switching of the material parameters---correctly modelling a dispersive time-varying material is not a trivial task.  The choice of model depends upon precisely how the time-variation is enacted within the material, not simply the initial and final values of the permittivity.  Ultimately, we have shown that when using these Drude-Lorentz models, the scattering coefficients will be heavily dependent on the initial conditions and upon the chosen model, and it may be possible to distinguish these models experimentally.

\begin{acknowledgments}
B.D. and C.M.H. acknowledge financial support from the Engineering and Physical Sciences Research Council (EPSRC) of the UK via the Exeter University Physics DTP. I.R.H. and S.A.R.H. acknowledge support from the EPSRC via the META4D Programme Grant (EP/Y015673/1).

\end{acknowledgments}

\bibliography{bib}

@article{kinsey2015,
  title={Epsilon-near-zero Al-doped ZnO for ultrafast switching at telecom wavelengths},
  author={Kinsey, N. and DeVault, C. and Kim, J. and Ferrera, M. and Shalaev, V. M. and Boltasseva, A.},
  journal={Optica},
  volume={2},
  number={7},
  pages={616--622},
  year={2015},
  publisher={Optical Society of America}
}

@article{tirole2022,
  title={Saturable time-varying mirror based on an epsilon-near-zero material},
  author={Tirole, R. and Galiffi, E. and Dranczewski, J. and Attavar, T. and Tilmann, B. and Wang, Y.-T. and Huidobro, P. A. and Al{\'u}, A. and Pendry, J. B. and Maier, S. A. and Vezzoli, S. and Sapienza, R.},
  journal={Physical Review Applied},
  volume={18},
  number={5},
  pages={054067},
  year={2022},
  publisher={APS}
}

@article{tirole2023,
  title={Double-slit time diffraction at optical frequencies},
  author={Tirole, R. and Vezzoli, S. and Galiffi, E. and Robertson, I. and Maurice, D. and Tilmann, B. and Maier, S. A. and Pendry, J. B. and Sapienza, R.},
  journal={Nature Physics},
  volume={19},
  number={7},
  pages={999--1002},
  year={2023},
  publisher={Nature Publishing Group UK London}
}

@article{Galiffi_2022,
   title={Photonics of time-varying media},
   volume={4},
   ISSN={2577-5421},
   url={http://dx.doi.org/10.1117/1.AP.4.1.014002},
   DOI={10.1117/1.ap.4.1.014002},
   number={01},
   journal={Advanced Photonics},
   publisher={SPIE-Intl Soc Optical Eng},
   author={Galiffi, Emanuele and Tirole, Romain and Yin, Shixiong and Li, Huanan and Vezzoli, Stefano and Huidobro, Paloma A. and Silveirinha, Mário G. and Sapienza, Riccardo and Alù, Andrea and Pendry, J. B.},
   year={2022},
   month=feb }

@ARTICLE{1124533,
  author={Morgenthaler, F.R.},
  journal={IRE Transactions on Microwave Theory and Techniques}, 
  title={Velocity Modulation of Electromagnetic Waves}, 
  year={1958},
  volume={6},
  number={2},
  pages={167-172},
  doi={10.1109/TMTT.1958.1124533}}

@article{J_T_Mendonça_2002,
doi = {10.1238/Physica.Regular.065a00160},
url = {https://dx.doi.org/10.1238/Physica.Regular.065a00160},
year = {2002},
month = {jan},
publisher = {},
volume = {65},
number = {2},
pages = {160},
author = {J T Mendonça and P K Shukla},
title = {Time Refraction and Time Reflection: Two Basic Concepts},
journal = {Physica Scripta}
}

@Article{Bacot2016,
author={Bacot, Vincent
and Labousse, Matthieu
and Eddi, Antonin
and Fink, Mathias
and Fort, Emmanuel},
title={Time reversal and holography with spacetime transformations},
journal={Nature Physics},
year={2016},
month={Oct},
day={01},
volume={12},
number={10},
pages={972-977},
issn={1745-2481},
doi={10.1038/nphys3810},
url={https://doi.org/10.1038/nphys3810}
}

@article{Moussa_2023,
   title={Observation of temporal reflection and broadband frequency translation at photonic time interfaces},
   volume={19},
   ISSN={1745-2481},
   url={http://dx.doi.org/10.1038/s41567-023-01975-y},
   DOI={10.1038/s41567-023-01975-y},
   number={6},
   journal={Nature Physics},
   publisher={Springer Science and Business Media LLC},
   author={Moussa, Hady and Xu, Gengyu and Yin, Shixiong and Galiffi, Emanuele and Ra’di, Younes and Alù, Andrea},
   year={2023},
   month=mar, pages={863–868} }

@Article{Zhou2020,
author={Zhou, Yiyu
and Alam, M. Zahirul
and Karimi, Mohammad
and Upham, Jeremy
and Reshef, Orad
and Liu, Cong
and Willner, Alan E.
and Boyd, Robert W.},
title={Broadband frequency translation through time refraction in an epsilon-near-zero material},
journal={Nature Communications},
year={2020},
month={May},
day={01},
volume={11},
number={1},
pages={2180},
issn={2041-1723},
doi={10.1038/s41467-020-15682-2},
url={https://doi.org/10.1038/s41467-020-15682-2}
}

@article{Bohn:21,
author = {Justus Bohn and Ting Shan Luk and Simon Horsley and Euan Hendry},
journal = {Optica},
number = {12},
pages = {1532--1537},
publisher = {Optica Publishing Group},
title = {Spatiotemporal refraction of light in an epsilon-near-zero indium tin oxide layer: frequency shifting effects arising from interfaces},
volume = {8},
month = {Dec},
year = {2021},
url = {https://opg.optica.org/optica/abstract.cfm?URI=optica-8-12-1532},
doi = {10.1364/OPTICA.436324},
}

@Article{Pacheco-Peña2020,
author={Pacheco-Pe{\~{n}}a, Victor
and Engheta, Nader},
title={Temporal aiming},
journal={Light: Science {\&} Applications},
year={2020},
month={Jul},
day={20},
volume={9},
number={1},
pages={129},
issn={2047-7538},
doi={10.1038/s41377-020-00360-1},
url={https://doi.org/10.1038/s41377-020-00360-1}
}

@ARTICLE{1138637,
  author={Holberg, D. and Kunz, K.},
  journal={IEEE Transactions on Antennas and Propagation}, 
  title={Parametric properties of fields in a slab of time-varying permittivity}, 
  year={1966},
  volume={14},
  number={2},
  pages={183-194},
  doi={10.1109/TAP.1966.1138637}}

@ARTICLE{8434236,
  author={Koutserimpas, Theodoros T. and Fleury, Romain},
  journal={IEEE Transactions on Antennas and Propagation}, 
  title={Electromagnetic Waves in a Time Periodic Medium With Step-Varying Refractive Index}, 
  year={2018},
  volume={66},
  number={10},
  pages={5300-5307},
  doi={10.1109/TAP.2018.2858200}}

@article{Koutserimpas:22,
author = {Theodoros T. Koutserimpas},
journal = {J. Opt. Soc. Am. B},
number = {2},
pages = {481--489},
publisher = {Optica Publishing Group},
title = {Parametric amplification interactions in time-periodic media: coupled waves theory},
volume = {39},
month = {Feb},
year = {2022},
url = {https://opg.optica.org/josab/abstract.cfm?URI=josab-39-2-481},
doi = {10.1364/JOSAB.445176},
}

@article{PhysRevB.104.214308,
  title = {Temporal equivalent of the Brewster angle},
  author = {Pacheco-Pe\~na, Victor and Engheta, Nader},
  journal = {Phys. Rev. B},
  volume = {104},
  issue = {21},
  pages = {214308},
  numpages = {6},
  year = {2021},
  month = {Dec},
  publisher = {American Physical Society},
  doi = {10.1103/PhysRevB.104.214308},
  url = {https://link.aps.org/doi/10.1103/PhysRevB.104.214308}
}

@Article{Sounas2017,
author={Sounas, Dimitrios L.
and Al{\`u}, Andrea},
title={Non-reciprocal photonics based on time modulation},
journal={Nature Photonics},
year={2017},
month={Dec},
day={01},
volume={11},
number={12},
pages={774-783},
issn={1749-4893},
doi={10.1038/s41566-017-0051-x},
url={https://doi.org/10.1038/s41566-017-0051-x}
}

@article{Pacheco-Pena:20,
author = {Victor Pacheco-Pe\~{n}a and Nader Engheta},
journal = {Optica},
number = {4},
pages = {323--331},
publisher = {Optica Publishing Group},
title = {Antireflection temporal coatings},
volume = {7},
month = {Apr},
year = {2020},
url = {https://opg.optica.org/optica/abstract.cfm?URI=optica-7-4-323},
doi = {10.1364/OPTICA.381175},
}

@article{Lustig:18,
author = {Eran Lustig and Yonatan Sharabi and Mordechai Segev},
journal = {Optica},
number = {11},
pages = {1390--1395},
publisher = {Optica Publishing Group},
title = {Topological aspects of photonic time crystals},
volume = {5},
month = {Nov},
year = {2018},
url = {https://opg.optica.org/optica/abstract.cfm?URI=optica-5-11-1390},
doi = {10.1364/OPTICA.5.001390},
}

@article{Xiao:14,
author = {Yuzhe Xiao and Drew N. Maywar and Govind P. Agrawal},
journal = {Opt. Lett.},
number = {3},
pages = {574--577},
publisher = {Optica Publishing Group},
title = {Reflection and transmission of electromagnetic waves at a temporal boundary},
volume = {39},
month = {Feb},
year = {2014},
url = {https://opg.optica.org/ol/abstract.cfm?URI=ol-39-3-574},
doi = {10.1364/OL.39.000574},
}

@misc{solís2021timevaryingmaterialspresencedispersion,
      title={Time-Varying Materials in Presence of Dispersion: Plane-Wave Propagation in a Lorentzian Medium with Temporal Discontinuity}, 
      author={Diego M. Solís and Raphael Kastner and Nader Engheta},
      year={2021},
      eprint={2103.06142},
      archivePrefix={arXiv},
      primaryClass={physics.optics},
      url={https://arxiv.org/abs/2103.06142}, 
}

@article{Koutserimpas:24,
author = {Theodoros T. Koutserimpas and Francesco Monticone},
journal = {Opt. Mater. Express},
number = {5},
pages = {1222--1236},
publisher = {Optica Publishing Group},
title = {Time-varying media, dispersion, and the principle of causality},
volume = {14},
month = {May},
year = {2024},
url = {https://opg.optica.org/ome/abstract.cfm?URI=ome-14-5-1222},
doi = {10.1364/OME.515957},
}

@article{Jaffray:22,
author = {Wallace Jaffray and Soham Saha and Vladimir M. Shalaev and Alexandra Boltasseva and Marcello Ferrera},
journal = {Adv. Opt. Photon.},
number = {2},
pages = {148--208},
publisher = {Optica Publishing Group},
title = {Transparent conducting oxides: from all-dielectric plasmonics to a new paradigm in integrated photonics},
volume = {14},
month = {Jun},
year = {2022},
url = {https://opg.optica.org/aop/abstract.cfm?URI=aop-14-2-148},
doi = {10.1364/AOP.448391},
}

@Article{Bohn2021,
author={Bohn, Justus
and Luk, Ting Shan
and Tollerton, Craig
and Hutchings, Sam W.
and Brener, Igal
and Horsley, Simon
and Barnes, William L.
and Hendry, Euan},
title={All-optical switching of an epsilon-near-zero plasmon resonance in indium tin oxide},
journal={Nature Communications},
year={2021},
month={Feb},
day={15},
volume={12},
number={1},
pages={1017},
issn={2041-1723},
doi={10.1038/s41467-021-21332-y},
url={https://doi.org/10.1038/s41467-021-21332-y}
}

@article{PhysRevLett.130.203803,
  title = {Eigenpulses of Dispersive Time-Varying Media},
  author = {Horsley, S. A. R. and Galiffi, E. and Wang, Y.-T.},
  journal = {Phys. Rev. Lett.},
  volume = {130},
  issue = {20},
  pages = {203803},
  numpages = {6},
  year = {2023},
  month = {May},
  publisher = {American Physical Society},
  doi = {10.1103/PhysRevLett.130.203803},
  url = {https://link.aps.org/doi/10.1103/PhysRevLett.130.203803}
}

@article{PhysRevA.111.033507,
  title = {Symmetry-protected lossless modes in dispersive time-varying media},
  author = {Hooper, Calvin M. and Capers, James R. and Hooper, Ian R. and Horsley, Simon A. R.},
  journal = {Phys. Rev. A},
  volume = {111},
  issue = {3},
  pages = {033507},
  numpages = {10},
  year = {2025},
  month = {Mar},
  publisher = {American Physical Society},
  doi = {10.1103/PhysRevA.111.033507},
  url = {https://link.aps.org/doi/10.1103/PhysRevA.111.033507}
}

@Article{Cho2020,
author={Cho, Choonlae
and Wen, Xinhua
and Park, Namkyoo
and Li, Jensen},
title={Digitally virtualized atoms for acoustic metamaterials},
journal={Nature Communications},
year={2020},
month={Jan},
day={14},
volume={11},
number={1},
pages={251},
issn={2041-1723},
doi={10.1038/s41467-019-14124-y},
url={https://doi.org/10.1038/s41467-019-14124-y}
}

@article{LustigSegalSahaBordoChowdhurySharabiFleischerBoltassevaCohenShalaevSegev+2023+2221+2230,
url = {https://doi.org/10.1515/nanoph-2023-0126},
title = {Time-refraction optics with single cycle modulation},
title = {},
author = {Eran Lustig and Ohad Segal and Soham Saha and Eliyahu Bordo and Sarah N. Chowdhury and Yonatan Sharabi and Avner Fleischer and Alexandra Boltasseva and Oren Cohen and Vladimir M. Shalaev and Mordechai Segev},
pages = {2221--2230},
volume = {12},
number = {12},
journal = {Nanophotonics},
doi = {doi:10.1515/nanoph-2023-0126},
year = {2023},
lastchecked = {2025-07-17}
}

\end{document}